\documentclass[aps,pre,groupedaddress,showpacs,twocolumn,preprintnumbers,
amsmath,amssymb,floatfix]{revtex4}
\usepackage{graphicx}
\usepackage{bm}
\begin{document}

\title{Radial Distribution Function of Rod-like Polyelelctrolytes}
\author{Roya Zandi}

\affiliation{Department of Chemistry and Biochemistry, UCLA,
Box 951569, Los Angeles, California, 90095-1569}
\author{Joseph Rudnick}
\bibliographystyle{apsrev}

\affiliation{Department of Physics and Astronomy, UCLA, Box 951547,
Los Angeles, CA 90095-1547}

\author{Ramin Golestanian}

\affiliation{Institute for Advanced Studies in Basic Sciences, Zanjan
45195-159, Iran\\
Institute for Studies in Theoretical Physics and Mathematics, P.O.
Box 19395-5531, Tehran, Iran}

\date{\today}

\begin{abstract}

We study the effect of electrostatic interactions on the
distribution function of the end-to-end distance of a single
polyelectrolyte chain in the rod-like limit. The extent to which
the radial distribution function of a polyelectrolyte is
reproduced by that of a wormlike chain with an adjusted effective
persistence length is investigated.  Strong evidence is found for
a universal scaling formula connecting the effective persistence
length of a polyelectrolyte with the strength of the electrostatic
interaction and the Debye screening length.

\end{abstract}


\maketitle

\section{Conformational statistics of a polyelectrolyte in the rod-like
limit: Background
and general considerations}

Because of their fundamental importance in many key biological
processes, electrostatically charged polymers, or polyelectrolytes
(PE's), have been the subject of intense research recently.  These
polymers often have stiff structures due to their Coulomb
self-repulsion, and this structural property lies at the heart of
their biological functionality.  For example, the cytoskeletal network
of actin filaments plays a crucial role in the recovery of eukaryotic
cell shape in the face of the stresses imposed by cell movement,
growth, and division \cite{Alberts,eichinger,janmey}.  Furthermore,
the storage properties and accessibility of DNA is known to be both
constrained and controlled by its stiffness \cite{sinden,zandi}.
Clearly, a complete study of the elastic properties of these charged
biopolymers, which is now facilitated by single filament imaging and
manipulation techniques, is crucial to an understanding of their role
in nature \cite{Chatenay}.

In the case of neutral polymers, the wormlike chain (WLC) model of
Kratky and Porod \cite{wlc} provides a powerful and convenient
characterization of flexibility through the persistence length, which
quantifies the correlations of unit tangent vectors parallel to the
chain.  Odijk, Skolnick, and Fixman (OSF), have applied this notion to
PE's through the introduction of an ``electrostatic'' persistence
length $\ell_e$ \cite{odijk,fixman}.  They derived the relationship
\begin{eqnarray}
\ell_e & = & \ell_{\rm OSF} \nonumber \\
& \equiv & \beta/4 \kappa^2,
\label{osf}
\end{eqnarray}
where $\kappa^{-1}$ is the Debye screening length (a measure of the
ionic strength of the solvent), and $\beta=\ell_{\rm B}/b^2$ is the
strength of the electrostatic interaction.  In the latter quantity
$\ell_{\rm B}=e^{2}/\epsilon k_{B}T$ is the Bjerrum length with
$\epsilon$ the dielectric constant of the ion-free solvent, and $b$
the average separation between neighboring charges along the PE.

The notion of electrostatic persistence length, however, is believed
to have shortcomings, as it replaces the many length scales in a PE by
a single effective persistence length.  The first note of caution is
due to Odijk himself, who pointed out that if one tries to make use of
the concept of an effective WLC approximation for a PE of finite
length $L$, the corresponding persistence length should be
$L$-dependent \cite{odijk}.  Barrat and Joanny later noted that, in
fact, each deformation mode of a PE with a given wavelength has a
different effective persistence length \cite{joanny}.  Moreover, the
wavevector dependence of the rigidity causes deviations from the
linear dependency $\ell_e \propto \beta$ \cite{joanny,ha}.

These observations---which point to the complications that may arise
if one insists on an effective WLC scheme for the characterization of
PE's---follow from the study of the averages $\langle R \rangle$ and
$\langle R^2 \rangle$, where ${\bf R}={\bf r}(L)-{\bf r}(0)$ is the
end-to-end distance of the PE. It is not at all clear that a
description based on the matching of first or second moments of the
end-to-end length distribution is equivalent to a comprehensive study
of conformational properties of a PE. An appealing alternative is a
study of the full radial distribution function of the end-to-end
distance of PE's, which provides an excellent gauge of the
conformation statistics of polymers in general \cite{de genne} and
stiff chains in particular \cite{frey}.  This quantity, which can be
measured experimentally through fluorescence microscopy, allows one to
determine whether or not the WLC model accurately reflects the
properties of the polymer segments.  If the model has been shown to
apply, then the fit of the distribution to the data directly yields
the persistence length and, hence, the elastic modulus of the polymer.

In this article, we calculate the end-to-end distribution function of
a charged inextensible chain in its rod-like limit.  This distribution
is then utilized to investigate the validity of a WLC with an adjusted
persistence length as a model for a PE. We find that when the Coulomb
interaction is only a perturbation to the mechanical stiffness of the
chain, or when Debye screening is sufficiently strong, the full radial
distribution of the PE can be collapsed onto that of a WLC. However,
in the case of sizable Coulomb interactions and when the screening
length is insufficiently short, we find that the distribution {\em
cannot} be collapsed onto that of any WLC. To quantify the deviations
from WLC behavior, we define an effective $\ell_p$ as the persistence
length of the WLC that is most likely to have the same end-to-end
distance as the PE chain.  We discover strong evidence of a universal
scaling function relating the electrostatic persistence length,
$\ell_e$, to the intrinsic persistence length, $\ell_{p0}$, the Debye
screening length, $\kappa$ and the strength of electrostatic
interaction, $\beta$.

Because of the inextensibility of the PE's under consideration
\cite{frey}, we adopt the Kratky and Porod WLC model to describe
the bending energy of the chain.  In this model, a polymer is
represented by a space curve ${\bf r}(s)$ as a function of the arc
length parameter $s$.  The total energy of the chain, which is the
sum of the intrinsic elasticity and the electrostatic energy can
be written as

\begin{equation}
\frac{\cal H}{k_B T} =\frac{\ell_{p0}}{2} \int_{0}^{L} d s \left(
\frac{d {\bf t}}{ds}\right)^{2} + \frac{\beta}{2}\int_{0}^{L} d s
d s^{\prime} \frac{e^{-\kappa |{\bf r}(s) - {\bf
r}(s^{\prime})|}}{|{\bf r}(s) - {\bf r}(s^{\prime})|},
\label{energy}
\end{equation}
\noindent where ${\bf t}(s)$ is the unit tangent vector, and
$\ell_{p0}$ is the intrinsic persistence length of the PE. We do not
take into account the fluctuations in the charges localized to the
chain and in the counterion system that can give rise to attractive
interactions leading to chain collapse \cite{golestan}.  Because the
chain is in its rod-like limit, excluded volume does not play a role.
The double integral on the right hand side of Eq.  (\ref{energy}) is
cut off when $|s-s^{\prime}|<b$, where $b$ is the spacing between
charges on the PE.  The quantity $b$ also represents the intrinsic
``coarse-graining'' of the effective Hamiltonian, in that structure on
a smaller length scale is not encompassed by the model embodied in Eq.
(\ref{energy}).  In addition, the value of $b$ represents a lower
limit on the magnitude of the intrinsic persistence length
$\ell_{p0}$, in that there is no physical meaning to a persistence
length that is exceeded by the smallest length scale in the system.

The end-to-end distribution function is defined as follows:
\begin{equation}
G({\bf r})=\langle\delta({\bf r}- {\bf R})\rangle,
               \label{dist}
\end{equation}
where ${\bf R}={\bf r}(L)-{\bf r}(0)$.  The average in (\ref{dist}) is
over an ensemble of PE chains.  The function $G({\bf r})$ is, then, the
probability that a given chain in the ensemble will have an end-to-end
distance equal to ${\bf r}$.  We make use of the procedure Wilhelm and
Frey have implemented to calculate the end-to-end distribution
function for inextensible neutral polymers \cite{frey}.  We focus on
Inextensible chains with $\ell_{p0} \sim L$ or $\ell_{p0} \ll L$.
In both cases, our study is limited to regions where the combination
of intrinsic stiffness and repulsive strength of the Coulomb
interaction keeps the chain in its rod-like limit \cite{note1}.  That
is, the combination of intrinsic persistence length and Coulomb
repulsion acts to keep the chain nearly \emph{straight} over its
entire length, so that its end-to-end distance does not greatly
deviate from its total arc length.  We parameterize the contour in
terms of the tangent field: ${\bf t}(s)=(a_{x}(s),a_{y}(s),1)/
\sqrt{1+a_{x}^{2}(s)+a_{y}^{2}(s)}$, and then retain only terms up to
second order in the $a$'s in the measure factor, $H$, and in the
argument of the delta-function in Eq.  (\ref{dist}), which is
rewritten in terms of its Fourier representation.  Expanding
$a_{x}(s)$ and $a_{y}(s)$ in a cosine series, as mandated by the
open-end boundary conditions, and making use of the relationship
\begin{equation}
{\bf r}(s) - {\bf r}(s^{\prime})=\int_{s}^{s^{\prime}}d x \;{\bf
t}(x),
\label{endtoend}
\end{equation}
we obtain
\begin{equation}
G(r)=N \int_{-\infty}^{\infty} \frac{d\omega }{ 2 \pi} \; e^{i\omega
(1-r/L)} \prod_{n=1}^{\infty} {\left(\frac{1 }{
\lambda_{n}+i\omega}\right)},
                 \label{integ}
\end{equation}
in which $N$ is a normalization constant. The
$\lambda_{n}$'s are eigenvalues of the matrix
\begin{equation}
{\bf T}=(n \pi)^{2}(\ell_{p0}/L) {\bf I} + \beta L {\bf E},
\label{Tdef}
\end{equation}
where ${\bf I}$ is the unity matrix and ${\bf E}$ is the electrostatic
energy matrix in the cosine basis set.  Contour integration then
yields
\begin{equation}
G(r)=N\sum_{n} f(n) e^{-\lambda_{n} (1-r/L)},
               \label{dist2}
\end{equation}
where
\begin{equation}
f(n)=\prod_{i \neq n}1/(\lambda_{i}-\lambda_{n}).
\label{f(n)}
\end{equation}

The large eigenvalues of the matrix ${\bf T}$ are dominated by the
diagonal terms $(n \pi)^{2}(\ell_{p0}/L)$.  In other words, the effect
of the electrostatic interaction is swamped by semiflexible energetics
at short length scales.  In the case of a neutral polymer with an
intrinsic persistence length that is comparable with the total length
of the polymer, the above series converges very rapidly, as noted in
Ref.  \cite{frey}.  When the chain is charged, and in particular, when
the stiffness is due predominantly to electrostatic effects, more
terms in the series must be preserved in order to obtain a stable
answer for $G(r)$.  In the calculations described here, we truncate
the matrix ${\bf T}$ at a size much greater than the number of terms
needed to obtain an accurate answer for the series of Eq.
(\ref{dist2}).  This is because a larger dimension of the truncated
matrix leads to more accurate values for the lower eigenvalues (which
participate in the sum), and a more accurate $f(n)$.  In practice, we
increased the dimension of the matrices until the effect (on the
end-to-end distribution function) of a further increase in the size of
${\bf T}$ by a factor of two was less than a part
in a thousand.  A restriction on the size of the matrix arises from
the requirement that the coarse graining length $b$ not exceed the
smallest wavelengths appearing in the cosine basis set.  If the length
of the PE is $L$, this means that the size, $N$, of the basis set
satisfies $N \le L/b$.  At no point in our calculations was this
inequality violated.

\begin{figure}
\includegraphics[height=1.9in]{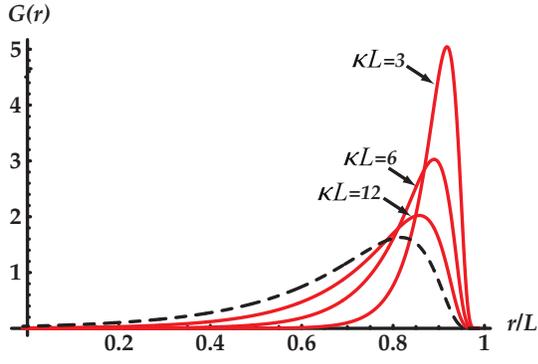} 
\caption{ End-to-end
distributions of charged semiflexible chains.  The thin solid lines
are plots of $G(r)$ for PE's with $\kappa L=12,6,3$ from left to
right.  In all cases $\beta L = 100$ and $\ell_{p0}/L=0.5$.  The
dashed line is the end-to-end distribution of a neutral semiflexible
chain with $\ell_{p0}/L=0.5.$ }
\label{fig:fig1}
\end{figure}

We set $b/L=10^{-3}$.  In this case, the distribution function of PE's
depends on three independent dimensionless parameters $\ell_{p0}/L$,
$\beta L$, and $\kappa L$.  Figure \ref{fig:fig1} illustrates the
effect of the screened electrostatic interaction on the distribution
function of a chain with $\ell_{p0}/L=0.5$ at different salt
concentrations and, hence, different values of $\kappa$.  As
illustrated in the figure, the end-to-end distance of the chain on
average becomes shorter upon a decrease of the screening length (for
fixed $\beta L$).  The quantity $\beta L$, which depends on the
solvent dielectric constant and the charge density on the PE chain, is
equal to 100 for all the three curves in the figure.

We now seek to determine under what conditions the end-to-end
distribution function of a rod-like PE is satisfactorily reproduced by
that of a WLC with an adjusted persistence length.  We also scrutinize
the adjusted persistence length of a WLC whose end-to-end distribution
is the closest fit to that of a rod-like PE to determine whether or
not this persistence length is in accord with existing predictions for
the dependence of the electrostatic persistence length on properties
of the PE chain.  In the cases we consider here the combination of intrinsic
stiffness---as parameterized by $l_{p0}/L$---and of charging---quantified
by $\beta L$---provides enough stiffening of the PE to guarantee
rod-like behavior in all length scales.

\section{Regimes in which the PE behaves like a WLC}

We find that there exist regimes in which the conformational
statistics of a PE chain in the rod-like limit are identical to those
of a WLC with an adjusted persistence length.  For such cases, PE and
WLC distributions are indistinguishable to the naked eye.  In these
regimes, the persistence length is as predicted by Eq.  (\ref{osf}),
or, when $\kappa L$ is not significantly larger than unity, by Eq.
(\ref{eqodijk}).  We are able to divide this regime into two different
categories.

\subsection{ Intrinsically stiff chains ($\ell_{p0} \sim L$)}
Whenever the electrostatic interaction plays a perturbative role in
the chain energetics, it is possible to obtain a satisfactory collapse
of the PE distribution onto that of a WLC. One noteworthy feature of
this regime is that the electrostatic persistence length always
satisfies $\ell_{e} < \ell_{p0}$.

\begin{figure}
\includegraphics[height=2in]{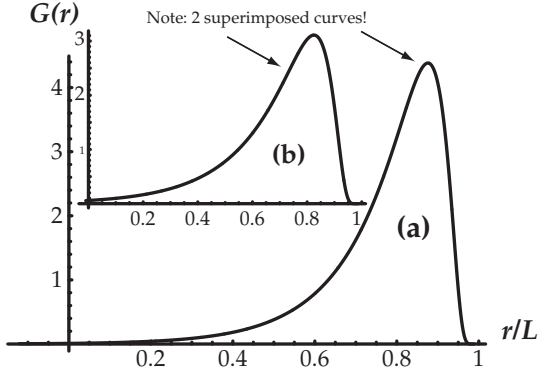}
\caption{The distribution for a polyelectrolyte with $\ell_{p0}=0.5$,
$\beta L=2400$, and $\kappa L=50$, compared to that of a wormlike
chain with $\ell_{p}=0.74$ (two superimposed curves and perfect match
with Eq. (\ref{osf})).  Inset: the distribution for a polyelectrolyte with
$\ell_{p0}=0.5$, $\beta L=5$, and $\kappa L=5$, compared to that of a
wormlike chain with $\ell_{p}=0.524$ (two superimposed curves and
perfect match with Eq.  (\ref{eqodijk})).  }
\label{fig:twocases}
\end{figure}

These regions are associated by weak charging of chain for small
values of $\kappa L$ ( $\kappa L \sim 1$ ).  As we increase screening,
$\kappa L$, the collapse of the two distributions can be achieved for
higher charge density.  Figure \ref{fig:twocases} illustrates two
examples of this regime.  In plot (a), the Debye screening length is
short, while $\beta L$, the strength of coupling, is high.  In plot
(b), the Debye screening length is relatively long, but the
electrostatic coupling is very weak.  In both cases, we find that the
distribution function of the PE's collapse onto WLC's with adjusted
persistence lengths that follows Odijk and OSF predictions (Eqs.
(\ref{eqodijk}) and (\ref{osf})) .

Our calculations verify that under physiological conditions ($\kappa=1
\ {\rm nm}^{-1}$), the distribution functions for rod-like DNA
segments ($L \raisebox{-0.03in}{$\stackrel{<}{\sim} $} 100 \ {\rm
nm}$) as well as those of stiffer actin filaments also collapse onto
the end-to-end distribution for neutral chains with an effective
persistence length given by Eq.  (\ref{osf}).

Whenever there is a virtually perfect collapse of the distribution
function of a PE onto that of a neutral chain, the persistence length
of the neutral chain follows Odijk's prediction, in that,
$\ell_{p}=\ell_{e}+\ell_{p0}$, where $\ell_{p}$ is the effective
persistence length of the charged chain, and $\ell_{e}=\ell_{\rm
Odijk}$ \cite{odijk}.
\begin{eqnarray}
\ell_{\rm Odijk}&=&\frac{\beta L^2}{12}\left[e^{-\kappa
L}\left(\frac{1}{\kappa L}+\frac{5}{(\kappa L)^{2}}+\frac{8}{(\kappa
L)^{3}}\right) \right.  \nonumber \\
&&\left.+\frac{3}{(\kappa L)^{2}}-\frac{8}{(\kappa L)^{3}}\right].
\label{eqodijk}
\end{eqnarray}
which reduces to $\ell_{OSF}\equiv \beta/4 \kappa^2$ for large $\kappa
L$ \cite{odijk,fixman}.  It is important to note that the expression
for $\ell_{\rm Odijk}$ \cite{odijk} was derived under the assumption
that the contour length of the chain, $L$, is of the order of its
intrinsic persistence length, $\ell_{p0}$ and that the electrostatic
interaction has the limited effect of only ``perturbing'' the WLC
shape of the chain \cite{odijk}.  Our results confirm the validity of
OSF and Odijk formulas in the regime in which they are expected to be
correct.

\subsection{Intrinsically flexible chains ($\ell_{p0} \ll L$)}

We are also able to identify regimes in which the radial distribution
function of a PE that is long compared to its intrinsic persistence
length ($\ell_{p0}/L \ll 1$) collapses almost perfectly onto that of
an uncharged WLC. Substantial charging is required to enforce the {\em
rod-like} limit for such chains.  It is important to note that despite
the substantial charging of the chain, the ratio of the length of the
PE to the screening length, $\kappa L$, must be sufficiently large.
This requirement is essential in order to minimize the strong
influence of the end effects on conformational statistics of charged
chains.

The important characteristic of this regime is that $\ell_{p0} \ll
\ell_{e}$ and thus electrostatic energy no longer plays a perturbative
role.  Nevertheless, if $\kappa L$ is sufficiently large we observe
collapse of the two distributions.  Furthermore, we also find that
Odijk's formula works to near perfection in those cases.  For example,
in the extreme instance of a very short intrinsic persistence length,
$\ell_{p0}/L=0.01$, a high degree of charging $\beta L=36000$, and a
screening length that is short compared to the chain length, $\kappa
L=100$, we obtain an end-to-end distribution that is nearly identical
with that of a WLC with an appropriate $\ell_{e}$.  In this case,
$\ell_{e}/L=0.894$ which is in near-perfect agreement with $\ell_{\rm
Odijk}$.  The accuracy of Odijk's formula when $\ell_{e} \gg
\ell_{p0}$ is not at all obvious, as OSF was derived in the regime
where electrostatic interaction plays a perturbative role.

We emphasize that the reason for the high quality of the match with
OSF in this regime is different from the reason for the corresponding
result obtained by Khokhlov and Khachaturian \cite{khokhlov} for
weakly charged flexible chains.  In our case, the chain is stiff in
all length scales; the possibility of renormalizing the length
and/or charge is thus excluded in our formulation.

\section{Regimes in which the conformational statistics of a PE
differ from those of a WLC}

\begin{figure}
\includegraphics[height=2in]{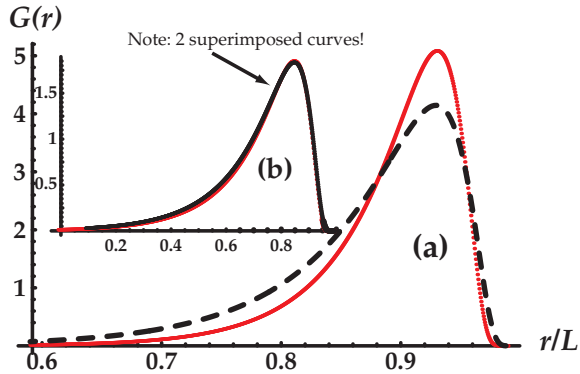}
\caption{The distribution for a polyelectrolyte with $\ell_{p0}=0.5$,
$\beta L=600$, and $\kappa L=10$, compared to that of a wormlike chain
with $\ell_{p}=1.3$.  Inset: the distribution for a polyelectrolyte
with $\ell_{p0}=0.5$, $\beta L=60$, and $\kappa L=10$, compared to
that of a wormlike chain with $\ell_{p}=0.6$ (two superimposed
curves).}
\label{fig:fig2}
\end{figure}

No matter how short the range of electrostatic interactions is as the
result of Debye screening, at sufficiently strong charging, the PE
end-to-end distribution differs significantly from that of a WLC.  The
emergence of a difference between the two distribution is accompanied
by a divergence between the effective persistence length
characterizing the conformational statistics and the predictions of
either Eq.  (\ref{osf}) or (\ref{eqodijk}).  Figure ~\ref{fig:fig2},
displays the PE end-to-end distribution (solid curves) along with the
modified WLC distribution (dashed curves) in a case in which it is
possible to obtain a nearly perfect fit (plot b) and in a case in
which the best fit not nearly as good (plot a).  In both cases fit was
obtained by matching the location of the maxima of the two
distributions, and the electrostatic persistence length attributed to
the PE distribution is that of the WLC associated with the dashed
curve.  The ratio of electrostatic persistence length extracted from
the distribution function to Odijk's persistence length is,
$\ell_e/\ell_{\rm Odijk}=0.99$ in the case of plot b.  For the inset
graph, plot a, in which the two distributions differ noticeably, this
ratio is $\ell_e/\ell_{\rm Odijk}=0.89$.

It is important to note that a rescaling of the backbone length of the
PE is not an acceptable stratagem for improving the agreement between
the PE radial distribution and that of a WLC. This is because the
backbone length is essentially fixed by the rod-like chain condition.
The shortening and thickening that is associated with intermediate
blob-like structures \cite{blobs} will not occur.  In fact, we have
observed that a reduction in the effective value of $L$ actually
degrades the quality of the correspondence between the conformational
statistics of a PE and those of the corresponding WLC.

Our general observation is that the ratio of $\ell_{e}$ extracted from
the effective WLC distribution to Odijk's persistence length
correlates with the quality of the fit of an effective WLC end-to-end
distribution to that of a PE. When this ratio is equal to one, the PE
is described to a high degree of accuracy in terms of a WLC. As this
ratio decreases, the deviation becomes more pronounced.  Figure
\ref{fig:fig3} displays a diagram which delineates the quality of the
fit.  The line that is used in the Figure to separate the two regimes
corresponds to $\ell_e/\ell_{\rm Odijk}=0.58$.  As indicated in this
Figure, for fixed $\kappa L$, when $\beta L$ is below a certain value
the PE behaves like a WLC, while for larger $\beta L$ there is a
substantial difference between the electrostatic persistence length of
a rod-like PE and Odijk's prediction.  As illustrated in
Fig.~\ref{fig:fig4}, the deviation of $\ell_{e}/L$ from $\ell_{\rm
Odijk}/L$, with increasing $\beta L$ is more pronounced at lower
values of $\kappa L$.  The Figure also shows that the electrostatic
persistence length of a rod-like PE, $\ell_{e}/L$, depends on
$\ell_{p0}/L$.  This dependence is not present in Odijk's formula.

\section{Universal Behavior}

\begin{figure}
\includegraphics[height=1.9in]{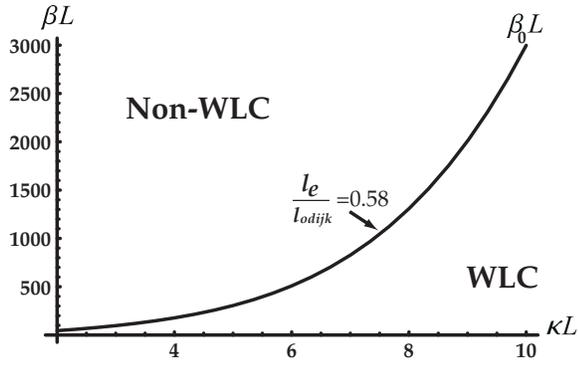} 
\caption{The diagram delineating the quality of a WLC fit for
$\ell_{p0}/L=0.5$.  The curve separating the two regimes corresponds
to $\ell_e/\ell_{\rm Odijk}=0.58$.  This line also indicates the
values of $\beta L$ in Eq.  (\ref{ours}) at different $\kappa L$'s.}
\label{fig:fig3}
\end{figure}

\begin{figure}
\includegraphics[height=1.9in]{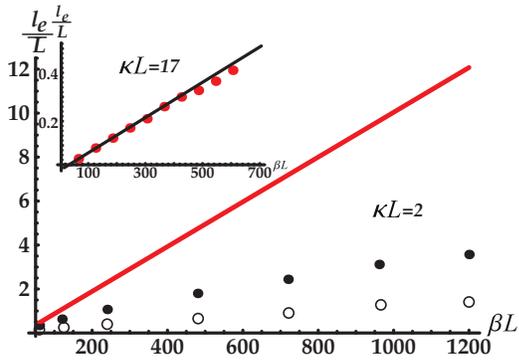}
\caption{Comparison of our results for the electrostatic persistence
length with Odijk's finite size formula \protect\cite{frey} (solid
line) at $\kappa L=2$ with $\ell_{p0}/L=0.5$ (filled circles) and
$\ell_{p0}/L=0.01$ (hollow circles).  The inset is for $\kappa L=17$
and $\ell_{p0}/L=0.5$.} \label{fig:fig4}
\end{figure}

\begin{figure}
\includegraphics[height=1.9in]{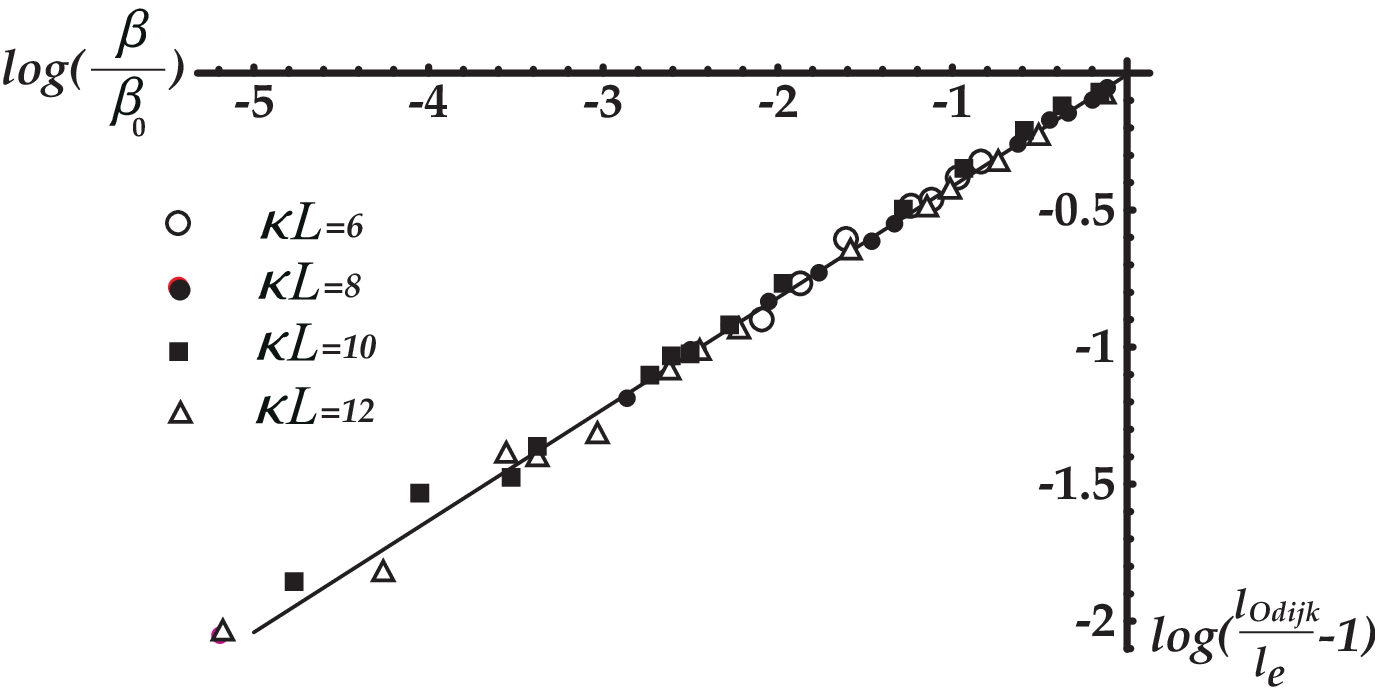}
\caption{Log-log plot of a comparison of our data, suitably rescaled,
to the expression in Eq.  (\ref{ours})
The points correspond to our rescaled plots of $\ell_{e}/L$ versus
$\beta L$ at $\kappa L=6,8,10,12$. } \label{fig:fig5}
\end{figure}

Our general observation of the curves of $\ell_{e}/L$ versus $\beta L$
is that they asymptote to a power law of the form $\ell_{e} \propto
(\beta L)^{1-x}$ where the exponent $x(\kappa L)$ lies between zero
and one.  In search of a possible universal behavior, which might
relate $\ell_{p0}/L$, $\beta L$ and $\kappa L$ to $\ell_{e}/L$, we
have rescaled our graphs for different values of the parameters.  The
details of rescaling will be presented in a forthcoming publication
\cite{rudnick}.  The dependence on charging, $\beta L$, of the ratio
of the persistence length of a WLC and the prediction for that
quantity embodied in Eq.  (\ref{eqodijk}) can be systematized in terms
of a universal formula\cite{itamar}.  This formula points to the
existence of a scaling mechanism.  At this point, we are not able to
provide an explanation for this mechanism or indicate a possible
underlying basis for it.  Figure~\ref{fig:fig5} contains plots of the
ratio of the electrostatic persistence length of PE distributions to
the predictions of Odijk in Eq.  (\ref{eqodijk}).  As shown in the
figure, all rescaled plots collapse on to a single curve corresponding
to the following crossover formula for $\ell_{e}$
\begin{equation}
\ell_{e}=\frac{\ell_{\rm Odijk}}{1+(\beta/\beta_{0})^{x}}
                \label{ours}
\end{equation}
This formula provides a remarkable fit to our data when the exponent
$x \simeq 0.4$, as exemplified in Fig.~ \ref{fig:fig5}.  The quantity
$\beta_{0}$ is an increasing function of $\ell_{p0}$ and $\kappa$.
Our most striking result is that for all investigated values of
$\kappa L$ and/or $\ell_{p0}/L$, the exponent $x$ remains fixed at
$\simeq 0.4$, independent of all the other parameters.  This function
encompasses all the regimes described above.  We are able to obtain
the value of $\beta_{0}$ from our data.  The crossover discussed in
the previous section occurs when $\beta \simeq 0.446 \beta_{0}$ which
corresponds to $\ell_{e} \simeq 0.58 \ell_{\rm Odijk} $ in Eq.
(\ref{ours}).  The curve in Fig.~ \ref{fig:fig3} indicates the values
of $\beta_{0} L$ at different $\kappa L$'s for $\ell_{p0}/L=0.5$.
Similar curves have been generated for other, smaller, values of
$\ell_{p0}/L$.  The behavior of these other curves is the same as the
one plotted in Fig.~\ref{fig:fig3}.

\section{Conclusions}

We conclude with a few brief comments.  To fit the PE and WLC distributions
we have matched the maxima of the two.  A fit can also be effected by
matching moments, or by a least-squares procedure.  Alternate fitting
approaches have been explored by us, but the results obtained were
entirely consistent with the conclusions set forth above.  The reason
for the present choice was the strong correlation between Odijk's
prediction and the fit of the distributions.  If we match the maxima
of the two distributions, the electrostatic persistence length departs
from Odijk's predictions when the distributions cease to closely
resemble each other, and thus the ratio of $\ell_e/\ell_{\rm Odijk}$
is a good measure of the similarity of the two distributions.
However, if we used a matching of the first or second moment to
calculate the effective persistence length, the two distributions are
clearly distinguishable from each other well within the regime in
which Odijk's formula for the persistence length remains accurate.

Our results indicate that the difference between the radial
distribution of the PE and the WLC can be attributed, at least in
part, to the influence of end effects.  In fact, we believe that the
behavior of the persistence length is substantially controlled by end
effects.  One way of understanding this is in terms of Odijk's
derivation of the expression (\ref{eqodijk}) for the electrostatic
persistence length \cite{odijk}.  This derivation is based on a
calculation of the energy of a bent segment of a charged rod.  A key
assumption in this derivation is that the segment takes the form of an
arc of a circle.  End effects are readily associated with the
difference between the shape of a real bent rod and the circular arc
assumed in Odijk's derivation.  An exploration of these effects in
this context will be described in forthcoming work \cite{shape}.

As noted above, the effect of counterion condensation has been
ignored throughout the above work.  It has been shown that counterion
condensation modifies the bending rigidity of a semiflexible chain
\cite{collapse,golestan} and may result in the collapse of the PE
chain \cite{golestan}.  We have performed a calculation of the
distribution function taking into account the attractive interaction
due to counterion fluctuations and observed the signature of collapse.
We are currently investigating these effects\cite{rudnick}.

In summary, we have been able to verify that the modified WLC as a
model of a rod-like PE works in the regime in which electrostatic
effects play a perturbative role, and that the Odijk and OSF forms for
the electrostatic persistence length are quantitatively accurate in
this regime.  We have also found that the above model and formulas are
accurate in non-perturbative regimes if Debye screening is
sufficiently strong.  In the regimes in which the above model fails,
we have discovered evidence for a scaling form, exhibited in Eq.
(\ref{ours}), for the correction to the Odijk result for the
persistence length.  As yet, no model provides a theoretical basis for
the scaling exhibited by this formula.

The authors would like to acknowledge helpful discussions with
W.M. Gelbart, M. Kardar, R.R. Netz, I. Borukhov, H. Diamant, K. -K.
Loh, V. Oganesyan, and G. Zocchi. This research was supported by
the National Science Foundation under Grant No. CHE99-88651.

\begin{references}
\bibitem{note1}
Numerical evaluation of $G(r)$ based on a Monte-Carlo simulation
in Ref. \cite{frey} reveals that the approximation taken for
analytical calculation of $G(r)$ is good when $L/\ell_{p}\leq 2$.
We have restricted our attention to this regime.
\end{references}
\end{document}